\documentclass[sn-chicago]{sn-jnl}
\jyear{2021}%
\usepackage[export]{adjustbox}
\usepackage[utf8]{inputenc} 
\usepackage[T1]{fontenc}    
\usepackage{url}            
\usepackage{booktabs}       
\usepackage{amsmath}
\usepackage{amssymb}
\usepackage{float}
\usepackage{epstopdf}
\usepackage{subcaption}
\usepackage[section]{placeins}
\usepackage{multirow}
\usepackage{bm}
\graphicspath{{/}}
\usepackage{pdflscape}
\usepackage{geometry}
\begin{document}

\title[Miscible viscous fingering in Hele-Shaw cells]{\vspace{-2.4cm}{The effect of viscosity ratio and Peclet number on miscible viscous fingering in a Hele-Shaw cell: A combined numerical and experimental study}}

\author[1]{\fnm{Daniel} \sur{Keable}}

\author[1]{\fnm{Alistair} \sur{Jones}}
\author[1]{\fnm{Samuel} \sur{Krevor}}
\author[1]{\fnm{Ann} \sur{Muggeridge}}
\author*[2]{\fnm{Samuel. J} \sur{Jackson}}\email{samuel.jackson@csiro.au}

\affil[1]{\orgdiv{Department of Earth Science \& Engineering}, \orgname{Imperial College London}, \orgaddress{\city{London}, \postcode{SW72BP}, \country{United Kingdom}}}

\affil*[2]{\orgdiv{CSIRO Energy}, \orgaddress{\city{Clayton South}, \state{Victoria}, \postcode{3169}, \country{Australia}}}

\abstract{{The results from a series of well characterised, unstable, miscible displacement experiments in a Hele Shaw cell with a quarter five-spot source-sink geometry are presented, with comparisons to detailed numerical simulation. We perform repeated experiments at adverse viscosity ratios from 1 – 20 and Peclet numbers from 10$^4$ – 10$^6$  capturing the transition from 2D to 3D radial fingering and experimental uncertainty. The open-access dataset provides time-lapse images of the fingering patterns, transient effluent profiles, and meta-information for use in model validation. 
\\
We find the complexity of the fingering pattern increases with viscosity ratio and Peclet number, and the onset of fingering is delayed compared to linear displacements, likely due to Taylor dispersion stabilisation. The transition from 2D to 3D fingering occurs at a critical Peclet number that is consistent with recent experiments in the literature. 2D numerical simulations with hydrodynamic dispersion and different mesh orientations provide good predictions of breakthrough times and sweep efficiency obtained at intermediate Peclet numbers across the range of viscosity ratios tested, generally within the experimental uncertainty. Specific finger wavelengths, tip shapes, and growth are hard to replicate; model predictions using velocity dependent longitudinal dispersion or simple molecular diffusion bound the fingering evolution seen in the experiments, but neither fully capture both fine-scale and macroscopic measures. In both cases simulations predict sharper fingers than the experiment. A weaker dispersion stabilisation seems necessary to capture the experimental fingering at high viscosity ratio, which may also require anisotropic components. 3D models with varying dispersion formulations should be explored in future developments to capture the full range of effects at high viscosity ratio and Peclet number.}}


\maketitle

\section{Introduction}
\label{sec_intro}

Viscous fingering is an instability that occurs when a less viscous fluid displaces a more viscous fluid in a porous medium or a Hele-Shaw cell. Small perturbations at the fluid-fluid interface (whether a sharp, immiscible interface, or a smooth miscible interface) grow, forming distinct fingers with characteristic wavelength and finger size. A similar form of fingering also results from gravitationally driven instabilities, occurring when a denser fluid sits above a lighter fluid (sometimes known as the Rayleigh-Taylor instability). The gravitationally driven instability was first described by \cite{Rayleigh1882} and further discussed by \cite{taylor1950} while the viscous driven instability was first investigated by \cite{Saffman1958} and is now often referred to as the Saffman-Taylor instability. Both types of instabilities have been studied extensively as they can influence the efficiency of carbon dioxide sequestration in the subsurface (e.g. \cite{Kopp2009, Ennis-King2005}), chromatographic experiments (e.g. \cite{Rousseaux2007}) and enhanced oil recovery (e.g. \cite{Gerritsen2005}). This paper focuses on viscous driven instabilities.

The fingering pattern structure for miscible, viscous driven instabilities is controlled by the viscosity ratio of the displaced to invading fluid ($M= \mu_2/\mu_1$), the transverse Peclet number ($Pe_T$)  and the anisotropy between longitudinal and transverse diffusion/ dispersion ($D_L/D_T$)  ( \cite{Tan1986, Zimmerman1991,Zimmerman1992, Hamid2020}) where the transverse Peclet number for linear displacements is defined as:
\begin{equation}
Pe_T = \frac{v L}{D_T}
\end{equation}
where $v$ is average fluid velocity, $L$ is a characteristic length and $D_T$ is the transverse diffusivity. For linear displacements (line drives) at a given rate, larger viscosity ratios result in an increased number of fingers with smaller characteristic wavelength, while higher transverse dispersivities result in fewer fingers. A higher longitudinal dispersion, for a given flow rate, will delay (or if very large, prevent) the onset of fingering by dispersing the interface between displacing and displaced fluids and making the displacement more stable. The ratio of longitudinal to transverse diffusivity controls the number of fingers and their growth rate (\cite{Tan1986, Zimmerman1991,Zimmerman1992, Hamid2020}). 

Miscible viscous fingering in radial flows is of practical relevance, as this may occur around injection wells in CO$_2$ storage schemes and during miscible gas injection for enhanced oil recovery. The fingering behaviours are more complex than in linear flows because the flow velocity decreases with distance from the injection point. This means that the impact of diffusion and dispersion on the fingering also changes with time and distance from the injector. This is further complicated because the dispersion may be velocity dependent at higher rates. In porous media the dispersion typically varies approximately linearly with velocity (\cite{Perkins1963}) whilst in Hele-Shaw cells the longitudinal dispersion varies with the square of velocity (\cite{Taylor1953a}) and the transverse dispersion is only dependent on molecular diffusion (\cite{Tan1987, Chen1998, Petitjeans1999}). In the absence of such velocity dependence, the Peclet number will decrease with distance from the injection point, so the number of fingers and their growth rate will also decrease (\cite{Tan1987}). For higher velocities in porous media, the impact of dispersion will tend to decrease with time until the flow is only affected by molecular diffusion (\cite{Riaz2004}). Some authors assert that the dependence of dispersion on velocity squared is not important (\cite{Petitjeans1999, Riaz2004}).

Numerical simulation is usually used to describe the impact of viscous fingering on systems of practical interest, however very high mesh resolutions are required to ensure that the solutions are dominated by physical diffusion and dispersion rather than numerical diffusion. This can make such simulations very computationally expensive, especially if a fixed mesh is used. A further limitation is that most finite volume methods, especially those using a Cartesian mesh, are affected by mesh orientation error when modelling miscible displacements, especially in radial flows (\cite{Brand1991}). Some researchers have explored the use of dynamic adaptive meshing to focus resolution in the fingered zone and thus reduce the computational effort required (e.g. \cite{Edwards1993, Lee2017, Kampitsis2020}). The mesh orientation effort is also much reduced in finite element or control-volume/finite element approaches using unstructured triangular meshes (e.g. \cite{Lee2017, Kampitsis2020}) or by using a nine-point method in finite volume methods such as described by \cite{Yanosik1979}). At larger scales, many researchers use upscaled, homogenized models (e.g. \cite{Koval1963, Todd1972}) to estimate the average impact on macroscopic flow, however, these models usually need to be calibrated by comparison with detailed simulation. Some insights can be obtained for early time behaviours from analytical solutions, usually in the form of linear stability analysis (e.g. \cite{Tan1986, Tan1987, Zimmerman1992, Riaz2004}) but these approaches are not suitable for predicting the late time, non-linear behaviours seen in most practical systems, especially when there is geological heterogeneity.

Physical experiments provide important insights into the physics of viscous fingering (as well as being essential in validating the predictions of detailed numerical simulations) however there are few experiments for radial flow reported in the literature. Not all of those that are reported provide sufficient information for validation purposes or they only investigate a small range of the possible physical behaviours. Ideally, there should be information regarding the values of all input parameters needed for simulation (system dimensions, fluid properties, porosity, permeability) as well as the experimental results in terms of volumes or rates of fluids produced over time and images of the displacement at different time intervals, including early times. 

Table \ref{tab_hele_shaw} and \ref{tab_porous_media} provides a summary of the experimental investigations of unstable miscible viscous fingering in radial flow for Hele-Shaw cells and porous media, respectively. We include references that provide sufficient information to characterise the pack or cell properties, the viscosity ratio(s) used and the Peclet number and/or provide some results in terms of fingering patterns or other measures of fingering dynamics. Most of these investigations used Hele-Shaw cells (\cite{Mahaffey1966,Paterson1985,Stoneberger1988, Petitjeans1999, Bischofberger2014, Djabbarov2016, Videbek2019}). \cite{Zhang1997} used glass bead packs whilst \cite{Habermann1960} used a system in which sand was glued to the glass plates. \cite{Lacey1961} performed experiments in glass beadpacks but did not provide data on their porosity and permeability. Nonetheless they are included in our review as they provide data on recovery efficiencies. The table also provides an estimate of the Peclet number, $Pe$ for radial flow (as defined by \cite{Petitjeans1999}): 

\begin{landscape}

\thispagestyle{empty}

\begin{table}[]
\hspace{-30pt}
\begin{tabular}{|c|c|c|c|c|c|c|c|c|}
\hline
\textbf{Hele-Shaw}                                                                               & \textbf{\begin{tabular}[c]{@{}c@{}}This \\ experiment\end{tabular}}     & \textbf{\begin{tabular}[c]{@{}c@{}}Mahaffey \\ $\bm{et}$ $\bm{al.}$ (1966)\end{tabular}}                         & \textbf{\begin{tabular}[c]{@{}c@{}}Paterson \\ (1985)\end{tabular}} & \textbf{\begin{tabular}[c]{@{}c@{}}Stoneberger \\ \& Claridge \\ (1988)\end{tabular}} & \textbf{\begin{tabular}[c]{@{}c@{}}Petitjeans \\ $\bm{et}$ $\bm{al.}$ (1999)\end{tabular}} & \textbf{\begin{tabular}[c]{@{}c@{}}Bischofberger \\ $\bm{et}$ $\bm{al.}$ (2014)\end{tabular}} & \textbf{\begin{tabular}[c]{@{}c@{}}Djabbarov\\ $\bm{et}$ $\bm{al.}$ \\ (2016)\end{tabular}} & \textbf{\begin{tabular}[c]{@{}c@{}}Videback \\ $\bm{et}$ $\bm{al.}$ (2019)\end{tabular}} \\ \hline
\textbf{Pattern}                                                                                 & \begin{tabular}[c]{@{}c@{}}Quarter \\ five spot\end{tabular}                & Five spot                                                                                          & Circle                                                              & \begin{tabular}[c]{@{}c@{}}Quarter \\ five spot\end{tabular}                          & \begin{tabular}[c]{@{}c@{}}Quarter \\ five spot\end{tabular}                 & Circle                                                                          & \begin{tabular}[c]{@{}c@{}}Quarter \\ five spot\end{tabular}                  & Circle                                                                     \\ \hline
\textbf{Outlet B.C.}                                                                             & \begin{tabular}[c]{@{}c@{}}Point sink \\ in opposite \\ corner\end{tabular} & \begin{tabular}[c]{@{}c@{}}Point sinks \\ in corners\end{tabular}                                  & \begin{tabular}[c]{@{}c@{}}Open \\ \end{tabular}       & \begin{tabular}[c]{@{}c@{}}Point sink \\ in opposite \\ corner\end{tabular}           & Point sink                                                                   & \begin{tabular}[c]{@{}c@{}}Open \\ \end{tabular}                   & \begin{tabular}[c]{@{}c@{}}Point sink in \\ opposite \\ corner\end{tabular}   & \begin{tabular}[c]{@{}c@{}}Open \\ \end{tabular}              \\ \hline
\textbf{\begin{tabular}[c]{@{}c@{}}System side \\ length or \\ diameter [cm]\end{tabular}} & 40                                                                          & 30.5                                                                                               & 60                                                         & 28.6                                                                                  & 68                                                                & 28                                                                              & 50                                                                            & 14                                                                      28 \\ \hline
\textbf{Plate spacing [cm]}                                                                    & 0.025                                                                       & \begin{tabular}[c]{@{}c@{}}0.00381,\\  0.0119\end{tabular}                                         & 0.15, 0.3                                                           & 0.01                                                                                  & 0.061                                                                        & \begin{tabular}[c]{@{}c@{}}0.0076, \\ 0.1143\end{tabular}                       & 0.0375                                                                        & \begin{tabular}[c]{@{}c@{}}0.0076, \\ 0.0419\end{tabular}                  \\ \hline
\textbf{Permeability  [D]}                                                                       & 5280                                                                       & 122.57, 1200                                                                                       & \begin{tabular}[c]{@{}c@{}}190000, \\ 760000\end{tabular}           & 844                                                                                   & 31400                                                                       & 488, 110000                                                                     & 11900                                                                        & 488, 14800                                                                 \\ \hline
\textbf{\begin{tabular}[c]{@{}c@{}}Flow rate \\ $[$ml/min$]$ \end{tabular}}                           & 0.1 - 10                                                                    & 0.049                                                                                              & 7.38, 73.80                                                         & 8.30                                                                                  & 0.0156 - 23.5                                                                & 0.4 - 40                                                                        & 1                                                                             & 0.001 - 10                                                                 \\ \hline
\textbf{\begin{tabular}[c]{@{}c@{}}Average \\ velocity [cm/s]\end{tabular}}                      & 0.0024 - 0.24                                                              & \begin{tabular}[c]{@{}c@{}}\textless{}0.02 (small gap)\\ \textless{}0.006 (large gap)\end{tabular} & \begin{tabular}[c]{@{}c@{}}0.0290 \\ - 0.580\end{tabular}           & 19.6                                                                                  & \begin{tabular}[c]{@{}c@{}}0.00603 \\ - 9.09\end{tabular}                    & 0.0443 - 0.29                                                                   & 0.0126                                                                        & \begin{tabular}[c]{@{}c@{}}0.00022 \\ - 0.040\end{tabular}                 \\ \hline
\textbf{\begin{tabular}[c]{@{}c@{}}Viscosity \\ ratio, $M$ \end{tabular}}                              & 1, 2, 5, 10, 20                                                             & 1, 3.3, 12.5, 39.4                                                                                 & \begin{tabular}[c]{@{}c@{}}Water and \\ glycerin\end{tabular}       & \begin{tabular}[c]{@{}c@{}}0.1, 2, 1.3,\\5.3, 50\end{tabular}                       & \begin{tabular}[c]{@{}c@{}}3.32, 4.48, \\ 12.2, 110,\\  153\end{tabular}     & 1 - 1350                                                                       & \begin{tabular}[c]{@{}c@{}}1, 2, 5, \\ 10, 20, 30, \\ 100\end{tabular}        & 52, 31, 909                                                                \\ \hline
\textbf{$\bm{Pe =Q/C \pi D_m h}$}                                                                             & 4240 - 424000                                                               & 1100 - 3400                                                                                        & \begin{tabular}[c]{@{}c@{}}6500 -\\131000\end{tabular}            & 881000                                                                                & 70 - 102000                                                                  & \begin{tabular}[c]{@{}c@{}}3100 - \\ 4650000\end{tabular}                       & 28300                                                                         & 6 - 350000                                                                 \\ \hline
\end{tabular}
\caption{\label{tab_hele_shaw}Comparison of Hele-Shaw experiments investigating unstable miscible displacements in radial flow from the literature. The average velocity is calculated from the time taken for a stable displacement to travel from injector to producer for the specified flow rate and cell dimensions. Open B.Cs have the entire cell circumference open to the atmosphere.}
\end{table}

\end{landscape}

\begin{table}[]
\hspace{-10pt}
\begin{tabular}{|c|c|c|c|c|}
\hline
\textbf{Porous media}                                                                & \textbf{\begin{tabular}[c]{@{}c@{}}This \\ experiment\end{tabular}}      & \textbf{\begin{tabular}[c]{@{}c@{}}Habermann \\ (1960)\end{tabular}}                                                               & \textbf{\begin{tabular}[c]{@{}c@{}}Lacey \\ $\bm{et}$ $\bm{al.}$ (1960)\end{tabular}}                    & \textbf{\begin{tabular}[c]{@{}c@{}}Zhang \\ $\bm{et}$ $\bm{al.}$ (1996)\end{tabular}}  \\ \hline
\textbf{Pattern}                                                                     & \begin{tabular}[c]{@{}c@{}}Confined \\ quarter\\  five spot\end{tabular} & \begin{tabular}[c]{@{}c@{}}Confined \\ quarter \\ five spot\end{tabular}                                                           & \begin{tabular}[c]{@{}c@{}}Confined and \\ unconfined \\ quarter \\ five spot\end{tabular} & \begin{tabular}[c]{@{}c@{}}Confined \\ quarter\\  five spot\end{tabular} \\ \hline
\textbf{\begin{tabular}[c]{@{}c@{}}System side \\ length [cm]\end{tabular}} & 40                                                                       & 11.4, 38.1                                                                                                                         & 22.5, 45, 90                                                                               & 40                                                                       \\ \hline
\textbf{\begin{tabular}[c]{@{}c@{}}Plate \\ spacing $[$cm$]$ \end{tabular}}               & 0.025                                                                    & 0.3175, 0.635                                                                                                                      & \begin{tabular}[c]{@{}c@{}}0.3,0.8, \\ 1.25, 7.62\end{tabular}                             & 1                                                                        \\ \hline
\textbf{Permeability [D]}                                                            & 5280                                                                    & 4.5, 12, 20                                                                                                                        & -                                                                                          & 5                                                                        \\ \hline
\textbf{Porosity (\%)}                                                               & 100                                                                      & 24.2, 31                                                                                                                           & -                                                                                          & 38                                                                       \\ \hline
\textbf{\begin{tabular}[c]{@{}c@{}}Approx \\ diameters \\ of beads [$\mu$m]\end{tabular}}   & -                                                                        & $\approx$53-600                                                                                                                       & -                                                                                          & $\approx$100                                                                \\ \hline
\textbf{\begin{tabular}[c]{@{}c@{}}Flow rate \\ $[$ml/min$]$ \end{tabular}}               & 0.1 - 10                                                                 & 0.07 - 30                                                                                                                          &                                                                                           -  & 0.83 - 3.33                                                             \\ \hline
\textbf{\begin{tabular}[c]{@{}c@{}}Average \\ velocity [cm/s]\end{tabular}}          & 0.0024 - 0.24                                                            & 0.00022 - 0.242                                                                                                                    & \begin{tabular}[c]{@{}c@{}}0.0006 \\ - 0.0210\end{tabular}                                 & 0.0011 - 0.0042                                                          \\ \hline
\textbf{\begin{tabular}[c]{@{}c@{}}Viscosity \\ ratio, $M$\end{tabular}}                  & 1, 2, 5, 10, 20                                                          & \begin{tabular}[c]{@{}c@{}}0.037, 0.222, 1, \\ 1.4, 2.4, 2.95, \\ 4.6, 12.9,17.3, \\ 23.4, 27.0, 38.2, \\ 71.5, 130.7\end{tabular} & 1, 10, 42                                                                                  & 4, 12, 25                                                                \\ \hline
\textbf{$\bm{Pe =Q/C \pi D_m h}$ }                                                                 & 4240 - 424000                                                            & 390 - 330000                                                                                                                       & -                                                                                          & 3000 - 12000                                                             \\ \hline
\end{tabular}
\caption{\label{tab_porous_media}Comparison of porous media experiments investigating unstable miscible displacements in radial flow from the literature. Velocity calculation as Table \ref{tab_hele_shaw}. Note, this experiment was performed in a Hele-Shaw cell but is shown here for comparison.}
\end{table}

\begin{equation}
Pe = \frac{Q}{C \pi D_m h}
\label{eqn_pe}
\end{equation}
where $Q$ is the flow rate (m$^3$/s), $D_m$ is the molecular diffusivity (m$^2$/s), $h$ is the thickness of the cell (m) and $C=1/2$ for flow in a quarter five spot well pattern and $C=2$ when the source is in the centre of a circle. We used the same value of diffusivity (3$\times$10$^{-10}$m$^2$/s), for water/glycerol \cite{DErrico2004}\footnote{Estimated for a 20\% volume fraction of glycerol using data in \cite{DErrico2004}. This is only approximate for all experiments except for those described in this paper, but it enables a rough comparison of Peclet numbers.}) in all cases, although this value will be slightly different for other fluid pairs or if there is velocity dependent dispersion. The value of $Pe$  is very large in all experiments indicating the dominance of viscous effects over diffusion, although it ranges over several orders of magnitude. 

All of the experiments investigated the flow patterns in homogeneous systems, with the exception of those described by \cite{Djabbarov2016}. In addition, of the experiments identified in the literature, only \cite{Zhang1997} and \cite{Djabbarov2016} compared the predictions of numerical simulations with their experimentally observed fingering patterns. In both papers the simulations did not provide a good prediction of the fingering patterns although they were able to provide reasonable estimates of effluent profiles.  

In this paper we describe a series of well-characterised, unstable, miscible displacement experiments in a glass Hele-Shaw cell with a uniform plate separation. The impact of viscosity ratio and Peclet number on the fingering patterns are investigated. Both the fingering patterns and the effluent profiles are compared against the predictions from numerical simulations for varying viscosity ratio for an intermediate Peclet number case. The resulting dataset represents an ideal validation case for numerical methods.

\section{Methodology}
\label{sec_method}

A series of miscible displacement experiments were performed in a Hele-Shaw cell \cite{Hele-Shaw1898}) and then simulated with a finite difference simulator that has previously been validated against a variety of linear, miscible displacement experiments (\cite{Christie1985, Christie1987, Christie1990, Davies1991}). 

\subsection{Physical experiments}
\label{sec_methods_exp}

The physical experiments were performed in a horizontally orientated, square Hele-Shaw cell, 0.4m on each side. The 15mm glass plates were separated by a 0.25mm thick Mylar insert which ran the circumference of the cell. They were mounted in an aluminium frame that kept them centralised and allowed the plates to be clamped together to ensure the cell was sealed around the edges. Fluid was injected through a 15mm port in the top glass plate using an ISCO pump at constant rate. Fluid was produced from a 15mm port on the opposite corner at constant atmospheric pressure, and collected in a sealed electronic scale unit. The displacement represented that seen in a quarter five spot injection-production pattern, as is typical in subsurface operations. The permeability, $k$, of the low Reynolds number system can be calculated using the depth-averaged Stokes flow through the parallel plates of spacing $h$ (\cite{Greenkorn1964}) as $k = h^2/12$. For our system, $k = 5.208 \times 10^{-9}$m$^2$ = 5280 D.

\begin{figure*}
\includegraphics[width=\textwidth]{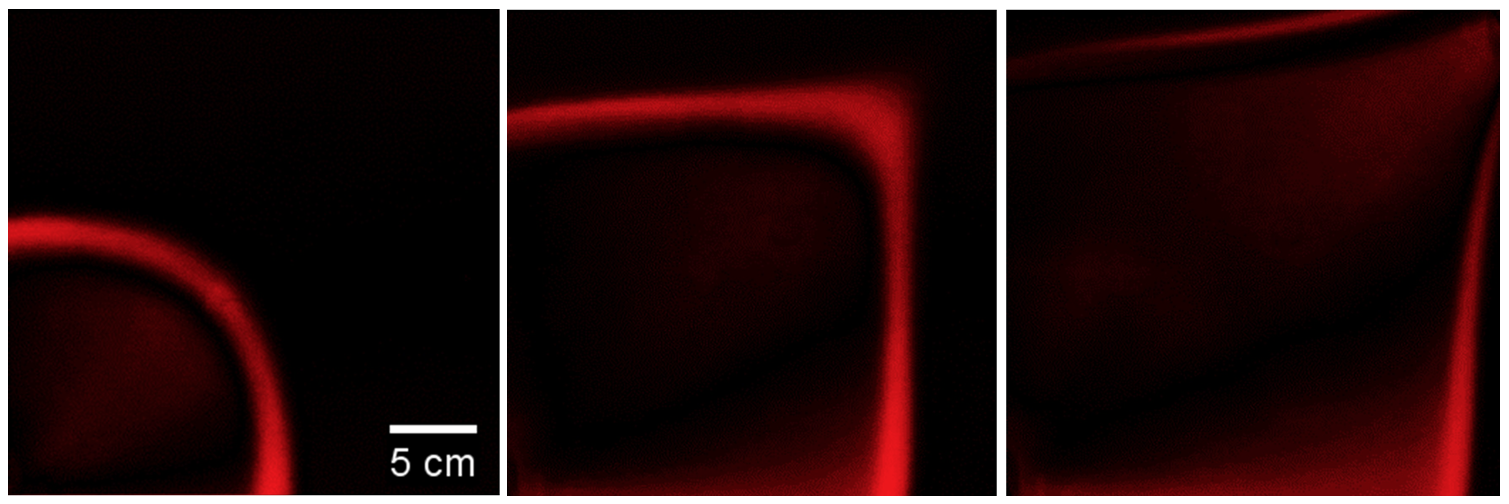}
\caption{The difference in flood front position, at 0.25, 0.5 and 0.75 PVI (left to right), for two $M$ = 1 stable displacements run in diﬀerent directions in the Hele Shaw cell. The images were rotated and overlaid with a difference filter to highlight any variances in the frontal advance (shown here in red). The red bands highlight the uncertainty in the frontal position between repeat, stable displacements}
\label{fig_repeatability} 
\end{figure*}

The uniformity of the gap and thus the homogeneity of the cell was confirmed by examining the interface observed during a stable (matched viscosity) displacement run in two diﬀerent directions (see description of the experimental fluids \& dyes below). The interface was symmetrical about the line joining the injection point to the outlet and similar in shape for both flow directions, indicated by the red band of uncertainty around the interface in (Figure \ref{fig_repeatability}). This uncertainty is typically less than the difference between fingering structures observed in different $M$, $Pe$ experiments.  

The fluids selected for the experiments were glycerol solution (displaced phase) and de-ionised water (DI, displacing fluid) since they can be used to achieve a range of viscosity ratios (by diluting the glycerol with water), have similar wetting properties on all materials used in the cell and are fully miscible. They have been used by several researchers for similar experiments including \cite{Habermann1960, Paterson1985, Zhang1997, Petitjeans1999}. Analytical expressions are available for estimating both the binary mixture density (\cite{Volk2018}) and viscosity (\cite{Cheng2008}) for different mixture proportions, making it straightforward to prepare the desired mobility ratios, (see Table \ref{tab_exp_mixtures}). The glycerol solution was dyed with Lissamine Green B at a concentration of 0.6g per 100ml. This provided enough colour contrast between the ﬂuids for imaging whilst being easy to clean from the cell and injection pump after finishing each experiment. The different concentrations used for the displaced fluid and associated viscosity ratios are given in Table \ref{tab_exp_mixtures}. These were determined using the relationship found in \cite{Cheng2008} at 20$^{\circ}$C under atmospheric conditions, and verified by direct measurements in several experiments, using a Brookﬁeld cone and plate rheometer at various shear rates.  The measured viscosity ratios were found to be within $\pm$10\% of those calculated, with differences largely due to temperature variations; these uncertainties are reflected in the horizontal error bars on the $M$ plots below. 

\begin{table}
\centering
\caption{The compositions of the displaced and displacing fluids used in the experiments with concentration of glycerol, $c_g$, expressed in \%volume, together with the viscosity ratios and Peclet numbers.}
\label{tab_exp_mixtures} 
\begin{tabular}{cccc}
\hline\noalign{\smallskip}
Displaced fluid & Displacing fluid & Viscosity Ratio, $M$ & Peclet number, $Pe$ \\
\noalign{\smallskip}\hline\noalign{\smallskip}
Mixture, $c_g$ = 20.3 \% & Mixture, $c_g$ = 20.3 \% & 1 & 1.4$\times$10$^5$\\
Mixture, $c_g$ = 20.3 \% & Water & 2 & 1.4$\times$10$^5$\\
Mixture, $c_g$ = 40.8 \% & Water & 5 & 1.4$\times$10$^5$\\
Mixture, $c_g$ = 53.0 \% & Water & 10 & 1.4$\times$10$^5$\\
Mixture, $c_g$ = 63.1 \% & Water & 20 &     \begin{tabular}[c]{@{}c@{}c@{}}1.4$\times$10$^4$, 7$\times$10$^4$ \\1.4$\times$10$^5$, 7$\times$10$^5$ \\  1.4$\times$10$^6$ \end{tabular}   \\
 \noalign{\smallskip}\hline
\end{tabular}
\end{table}
Images of the ﬂuid displacement were recorded using a camera mounted above the cell and time-lapse software. These images were post-processed to estimate breakthrough times for comparison with the measurements of produced fluids, as well as metrics such as interfacial lengths. They were first cropped and processed to account for small misalignments with the camera objective, using a projective geometric transformation in Matlab. The image brightness was scaled using pre and post-experiment images to account for uneven illumination across the imaging period. 

Images were segmented using the FIJI image processing software \cite{Schindelin2012}.  We utilised a ﬂat ﬁeld correction (FFC) to correct for uneven illumination or dirt/dust on the lenses using the pre-injection reference image. The image segmentation steps are then: conversion from RGB space to 16-bit greyscale, FFC, brightness and contrast correction (in some cases), application of a sharpening mask followed by auto-thresholding based on the RenyiEntropy algorithm (Kapur et al., 1985). The auto-threshold limits were manually checked, with resulting metrics largely independent of threshold choice (e.g. variations of $\pm$5\% generated $<$2\% change in areal sweep calculations).

\subsection{Numerical modelling}
\label{sec_methods_mod}

The experiments were modelled using a ﬁnite-diﬀerence numerical method based on that first described by \cite{Christie1987}. This was originally developed to model unstable, first contact, miscible displacements and has been subsequently developed to include various higher order methods, velocity dependent dispersion where the dispersion depends on a specified power of the velocity and inactive cells, amongst other features. It has been previously validated against a range of physical experiments investigating linear displacements in porous media (see for example (\cite{Christie1987, Christie1990, Davies1991}), but has not previously been fully validated for radial flows or flows in Hele Shaw cells. 

For first contact miscible fluids we solve the following mass conservation equation:
\begin{equation}
\phi \frac{ \partial c}{\partial t} + \bm{v} \cdot \nabla c = \phi \nabla \cdot \left( \bm{D} \nabla c \right)
\label{eqn_mass_con}
\end{equation}
where $c$ is the mass fraction of the injected fluid, $\phi$ is the porosity (=1 in a Hele-Shaw cell), $\bm{D}$ is the dispersion tensor, $\nabla$ is the vector differential operator and $\bm{v}$ is the Darcy velocity given by:
\begin{equation}
\bm{v} = - \frac{k}{\mu_m}\nabla P
\label{eqn_darcy_v}
\end{equation}
where $k$ is the permeability of the porous medium, $\mu_m(c)$ is the mixture viscosity and $P$ is the fluid pressure. The exponential mixing rule (as suggested by \cite{Cheng2008}) was used to describe the mixture viscosity:
\begin{equation}
\mu_m = \mu_1 e^{-c \ln M}
\end{equation}
where $M = \mu_2 / \mu_1$, subscript 1 refers to the injected fluid, and subscript 2 to the displaced fluid. The dispersion tensor, $\bm{D}$ in Equation (\ref{eqn_mass_con}), is given by:
\begin{equation}
\bm{D} = \begin{bmatrix} D_L & 0\\0&D_T \end{bmatrix} 
\label{eqn_dispersion_tensor}
\end{equation}
where $D_T$ and $D_L$ are the the transverse and longitudinal dispersion coefficient, respectively. In this work, we set $D_T$ equal to the molecular diffusivity in the bulk fluid, $D_m$. The tensor is orientated with longitudinal in the direction of flow and transverse perpendicular to this, at all times. $D_L$ is given by:
\begin{equation}
\frac{D_L}{D_m} = 1 + \frac{2}{105}\frac{h^2 \lvert \bm{v}  \rvert^2}{D_m^2}
\label{eqn_dispersion}
\end{equation}
where $\lvert \bm{v}  \rvert$ is the local speed. Equations (\ref{eqn_mass_con}) and (\ref{eqn_darcy_v}) are constrained by the equation for the conservation of total volume (as the system is assumed to be incompressible):
\begin{equation}
\nabla \cdot \bm{v} = 0
\label{eqn_con_vol}
\end{equation}
Combining Equation (\ref{eqn_con_vol}) with (\ref{eqn_darcy_v}), an elliptic pressure equation is formed:
\begin{equation}
\nabla \cdot \left( \left( \frac{k}{\mu_m} \right) \nabla P \right) = 0
\label{eqn_elliptic_p}
\end{equation}

To solve the system of equations numerically we employ an implicit pressure, explicit saturation (IMPES) technique with finite-differences, solving the elliptic pressure Equation (\ref{eqn_elliptic_p}) implicitly, with a first order backward Euler scheme, and spatially with a second order accurate central scheme. The solution technique uses a modified incomplete Cholesky (MIC(0)) preconditioner with Cholesky conjugate gradient method. The hyperbolic concentration transport Equation (\ref{eqn_mass_con}) is then solved explicitly using a flux-corrected transport algorithm (FCT) by \cite{Zalesak1979}. It uses uses operator splitting to include diffusion and dispersion which is solved implicitly. 

A dimensionless Cartesian mesh of size 400$^2$ was used for all 2D simulations with the mesh orientated parallel, or diagonal, to the line joining injection and production wells. This mesh resolution was chosen following a mesh refinement study to ensure that physical diffusion dominated over numerical diffusion. Fingers were triggered by specifying a log-normal permeability distribution with a standard deviation equivalent to $\pm$0.03mm and a correlation length of 2mm across the entire mesh. Five different realisations (with the same statistics) were simulated for the $M=20$ displacement in order to estimate the influence of different permeability distributions on the simulated behaviour. 

A constant rate injector and producer were specified in opposite corners of the square of active cells to act as a source and sink of fluid. These had the same radius as the inlet and outlet ports and locations with respect to the model boundaries as used in the experiments.

\section{Results}
\label{sec_results}

\subsection{Varying viscosity ratio, $M$}
\label{sec_results_varyingM}

\begin{figure}[]
\hspace{-30pt}
\includegraphics[width=1.2\textwidth]{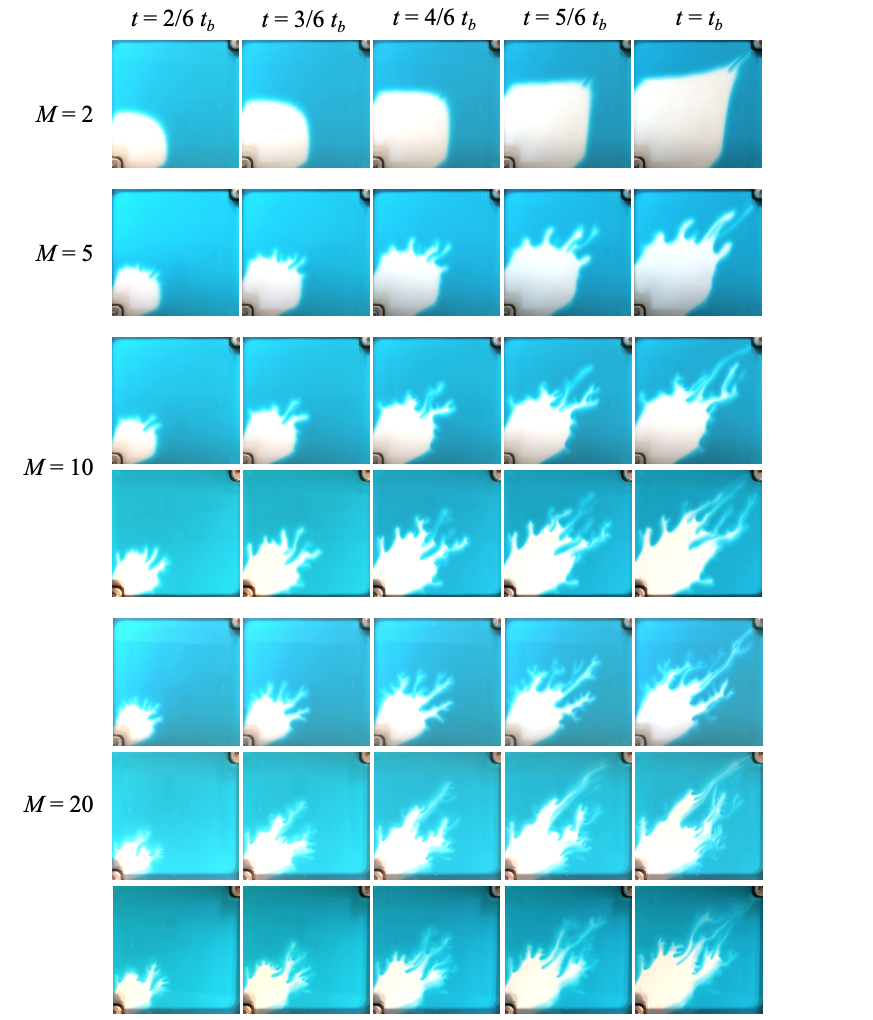}
\caption{Evolution of miscible fingering patterns over time, $t$ (expressed as fractions of breakthrough time, $t_b$, left to right) seen in experiments with different viscosity ratios ($M$, increasing top to bottom), at $Pe= 1.4 \times 10^5$. For $M=10, 20$ we show repeat experiments with source and sink swapped (images rotated) under the top images ($M=20$ has two repeats in the opposite direction). Note the bottom two $M=20$ repeat experiments have some shadow artefacts on the bottom of the cell, but are shown for qualitative comparisons.}
\label{fig_exp_varyingM} 
\end{figure}

\begin{figure}[ht!]
\centering
\includegraphics[width=1\textwidth]{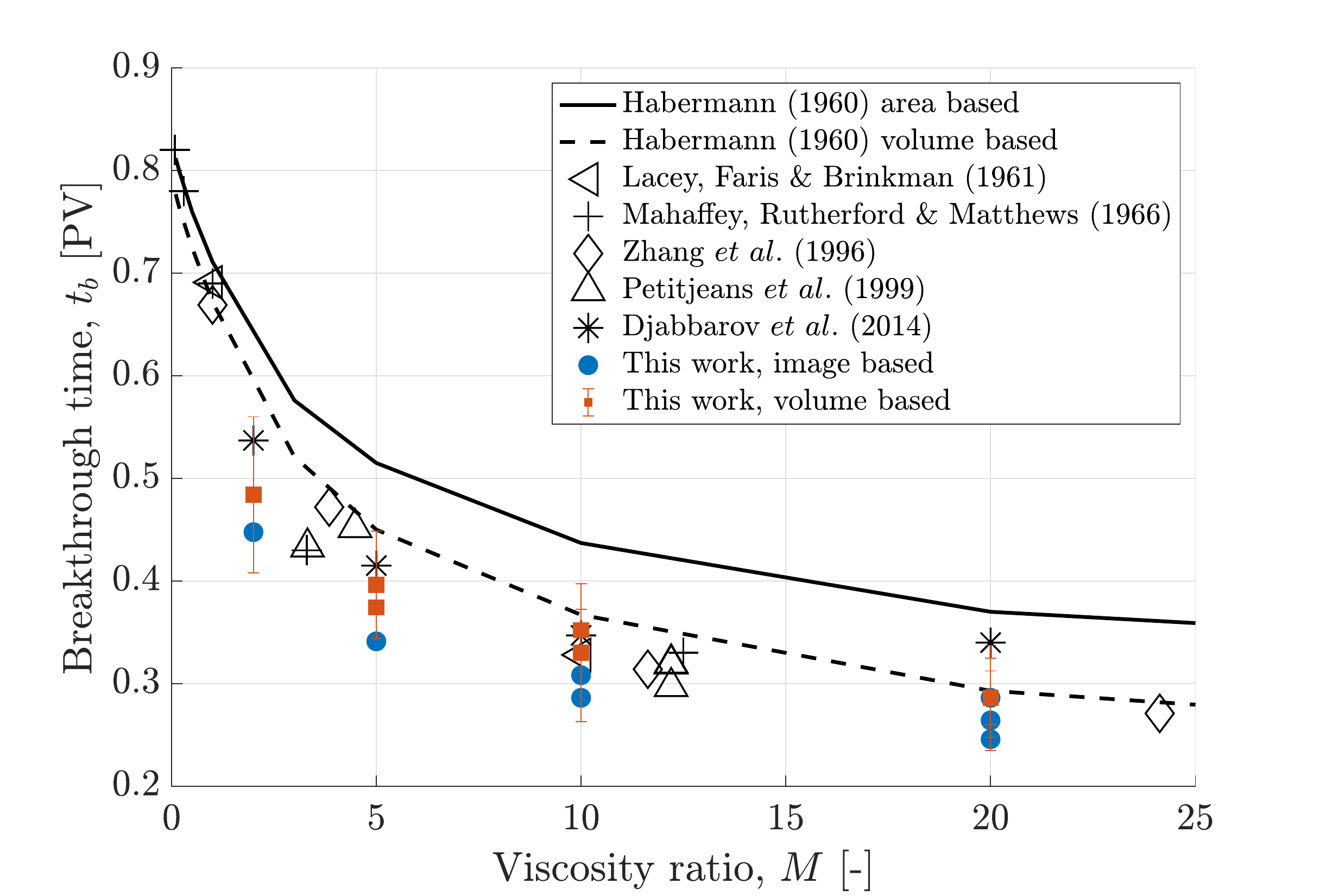}
\caption{Comparison of breakthrough times (in pore-volumes injected) obtained from the Hele Shaw experiments ($Pe= 1.4 \times 10^5$) with those reported in the literature for different viscosity ratios. Literature values are often reported as swept volume at breakthrough, which is equivalent. Circles show image based results, squares show volume based results. Multiple points at the same viscosity ratio represent repeat experiments. Note 3 of the red square points at $M = 20$ overlay a blue circle with the same breakthrough time. }
\label{fig_exp_bt_M} 
\end{figure}

Figure \ref{fig_exp_varyingM} shows the evolution of the fingering pattern in the experiment for different viscosity ratios where time is expressed in fractions of the breakthrough time, $t_b$, for that particular viscosity ratio ($Pe= 1.4 \times 10^5$ in all cases). We show the breakthrough time in pore-volumes injected for the varying viscosity ratio cases quantitatively in Figure \ref{fig_exp_bt_M}, along with other literature results. Experimental results are also shown for all cases in Table \ref{tab_exp_results} in Appendix A. The volume based breakthrough time is determined from the point at which the recovered vs. injected volume gradient $dV_{r,2}/dV_{i,1}$ inflects below unity. The image based breakthrough times are calculated as the time at which the displacing plume (in white) passes into the sink outlet. There is an uncertainty of $\pm 1$ frame with this method, with an error bar within the symbols in Figure \ref{fig_exp_bt_M}.

In-line with previous experimental results, we find that that breakthrough time is reduced as the viscosity ratio increases. This can be shown formally through linear stability analyses (\cite{Tan1987, Riaz2004}), as the number and the length of the fingers increase with viscosity ratio, resulting in an earlier breakthrough. As the viscosity ratio $M \rightarrow \infty$, the breakthrough time plateaus. There is increasing variability in finger velocities at higher $M$, with more shielding of small finger growth (\cite{Nicolaides2015}). This leads to a more dominant single finger, along the central source-sink streamline (see Figure \ref{fig_exp_varyingM} bottom right), which largely controls breakthrough and becomes invariant to $M$ as $M$ becomes very large. 

The experimental results in Figure \ref{fig_exp_varyingM} and \ref{fig_exp_bt_M} show the variability in repeated experiments, along with sources of experimental uncertainty. This is largely lacking in previous experimental literature, but provides realistic bounds for assessing model conformance.  Experiments were repeated in opposite flow directions in the cell, swapping the source and sink (for $M$=5,10,20) as well repeated in the same direction (for $M$=20) to highlight typical variability. Figure \ref{fig_exp_varyingM} shows that similar fingering structures, with characteristic wavelength, are created in repeat runs, but the specific finger growth, and tip-splitting occurrence is distinct in each case. This again highlights the cell homogeneity - there are no inclusions or regions that specifically trigger instabilities. Instead, fingering variations between repeat runs are likely due to small instabilities in the initial fluid configuration as it leaves the injection port and the interaction thereafter with the cell varying thickness. Small variations in the cell thickness create variations in spatial permeability; this is how instability is triggered in the simulations and represents a realistic finger formation mechanism. The stochastic nature is captured in the simulations by varying the statistical realization of the permeability field, mimicking the experimental uncertainty (see Figure \ref{fig_exp_sim_BT_rec} below).

The volume based experimental results in Figure \ref{fig_exp_bt_M} have uncertainty bars which stem from the difference in recovered volume of displaced fluid (estimated from mass produced and calculated mixture density), and the injected volume of displacing fluid (estimated from constant pump volume rate) at the breakthrough point - they should be equal if there were no uncertainty.  Generally the image based breakthrough is slightly faster than the volume based method, largely due to the precision of the volume based method which has a detection like 'threshold' for when the recovered volume has clearly diverged from the injection volume. The correspondence of the image and volume based methods (when considering the uncertainty) gives confidence in the experimental precision, and the accuracy of the results when compared to other literature values across a range of conditions. The image based method is more reliable than the volume based methods when the fingering is truly 2D (i.e not at high $Pe$ in further results), however it does not give information after breakthrough. Generally the repeat breakthrough times vary by $\approx$0.02PV through the image based methods, which represents variation of 5-8\% of the breakthrough time over the range considered here. This variability is smaller than the volume based uncertainty in most cases. 

\begin{figure}
\centering
\includegraphics[width=\textwidth]{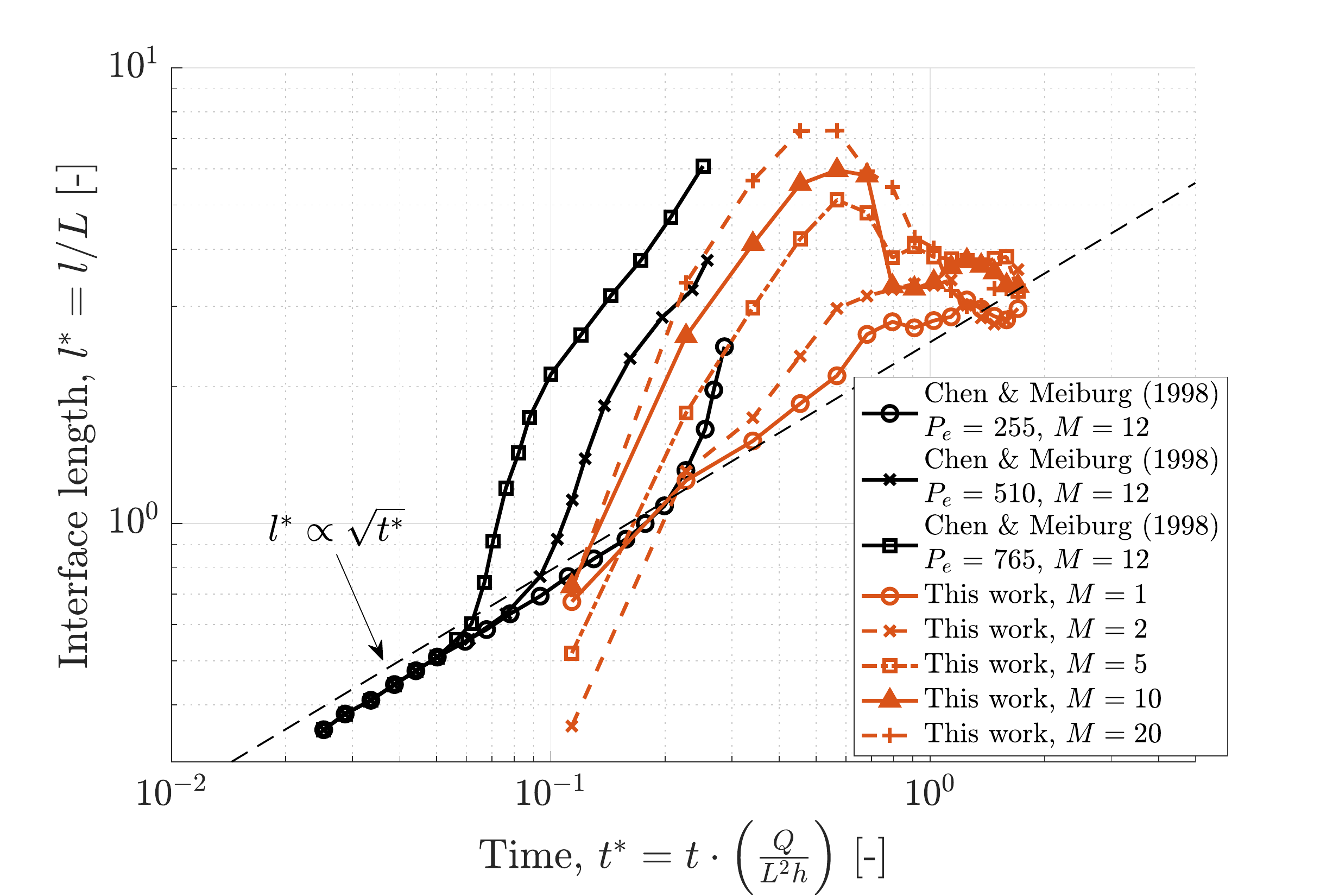}
\caption{Dimensionless interface length, $l^*$, with dimensionless time $t^*$ for different viscosity ratios ($Pe= 1.4 \times 10^5$) compared with simualtion data from \cite{Chen1998}. The dashed line shows the $\sqrt{t^*}$ scaling of $l^*$ for a stable growing interface (note, the line is shown for the gradient, but can be offset by a constant for different configurations). The interface length increases with mobility ratio due to the increased number and length of fingers until breakthrough. After breakthrough the displacing fluid channels along the finger closest to the diagonal joining injector to producer and subsidiary fingers grow more slowly so the interface length decreases}
\label{fig_exp_int_length} 
\end{figure}

In Figure \ref{fig_exp_int_length} we show the dimensionless interface length against time for the different viscosity ratio experiments. The interface length was calculated directly from the segmented images using the perimeter of the displacing phase (this is generally smooth so there are no finite size effects). The simulation results from \cite{Chen1998} for M= 12, $Pe = 255, 510$ and 765 (converted to the $Pe$ definition here, their $Pe$ values are 400, 800 and 1200, respectively) are shown for comparison. In all cases the interface length increases with time to a maximum at breakthrough. After breakthrough the interface length decreases again. This is mainly due to the decay of side fingers as the displacing fluid channels preferentially along the finger(s) that have broken through. We also show the theoretical scaling of a stable growing circle subject to mass conservation ($l^* \sim \sqrt{t^*}$), which the $M$ = 1 case follows for its displacement until breakthrough. The \cite{Chen1998} results are qualitatively similar but the interface length increases more rapidly than in our experiments. This is despite their much lower Peclet number which would suggest their flows are more influenced by diffusion and should thus have fewer fingers growing more slowly. 

\begin{figure}
\includegraphics[width=\textwidth]{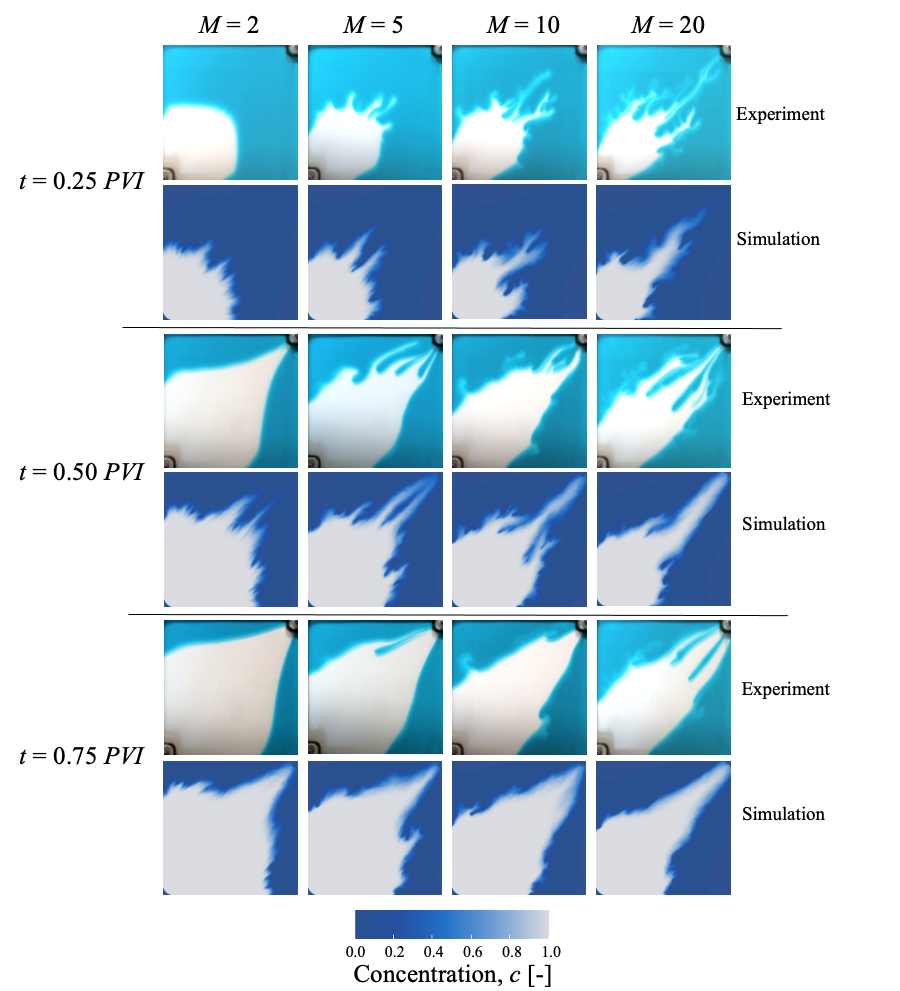}
\caption{Comparison of fingering patterns predicted by simulation with experimental observations for $M= 2  - 20$ ($Pe= 1.4 \times 10^5$) with a parallel mesh at different pore-volume injected times. Simulation concentration legend is shown at the bottom of the figure.}
\label{fig_exp_sim_M} 
\end{figure}

In our experiment results, we see a significant delay in the onset of fingering in all cases (see Figure \ref{fig_exp_varyingM}). This has been observed and discussed previously in linear displacements  (see \cite{Hamid2020} and references therein). It was also seen for radial flows in the experiments reported by \cite{Perkins1963, Paterson1985, Videbek2019}  although only discussed by \cite{Perkins1963}. The phenomenon is also discussed briefly in the analytical study by \cite{Sharma2019}. This delay occurs because initially the step interface between displaced and displacing fluid spreads more rapidly by diffusion/dispersion than the fingers can grow. This also reduces the initial number of fingers that form and their subsequent growth rate (see for example \cite{Perkins1963}). We hypothesize that this initially stable zone is larger in radial flows subject to velocity dependent dispersion because of increased velocities near the injector. This may be exacerbated in Hele-Shaw cells compared with porous media because the dispersion is dependent on the square of velocity, see Equation (\ref{eqn_dispersion}). This may also explain why the interface length obtained from our experiments (Figure \ref{fig_exp_int_length}) grows more slowly than the \cite{Chen1998} simulations. Their simulations neglected hydrodynamic dispersion and were performed using much lower Peclet numbers. We estimate that velocity dependent dispersion dominates over molecular diffusion in our experiments for the duration of each displacement. 

\begin{figure}[ht!]
\hspace{-35pt}
\subfloat[]{\includegraphics[width=0.65\textwidth]{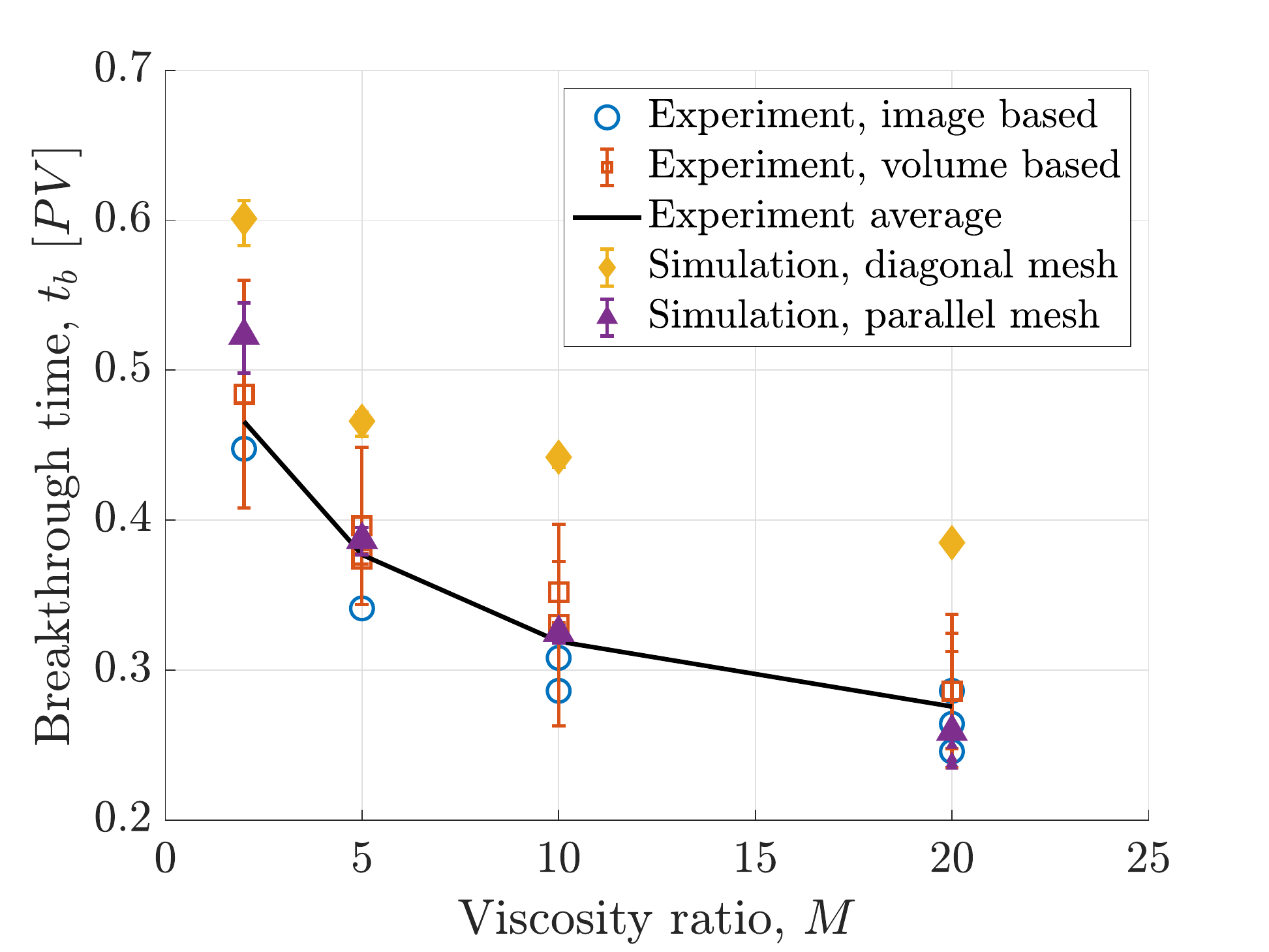}}
\hspace{-20pt}
\subfloat[]{\includegraphics[width=0.65\textwidth]{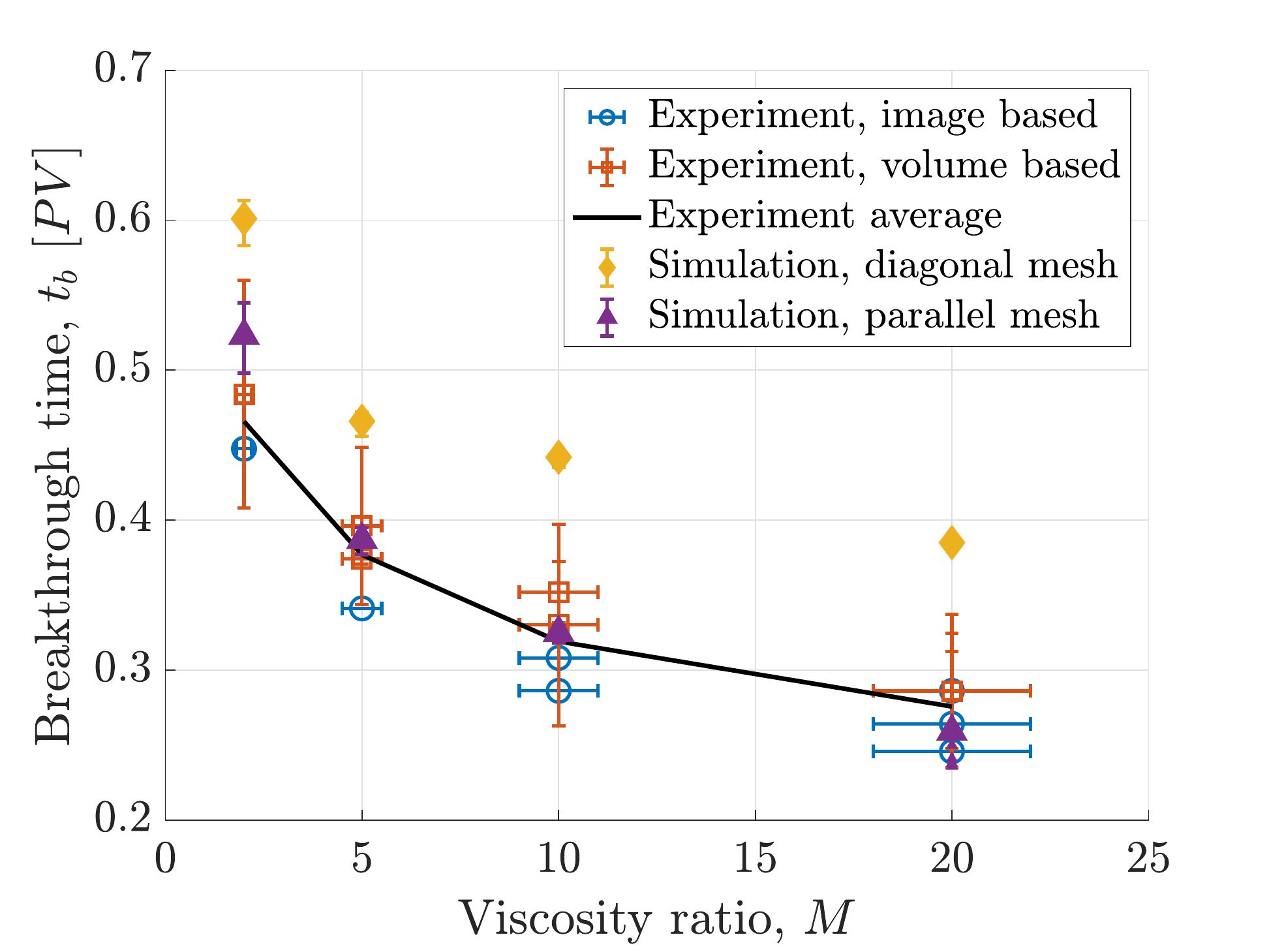}}
\caption{Numerical simulation predictions of (a) breakthrough time (in pore-volumes, $PV$) and (b) recovery after 1 pore-volume injected $PVI$, compared to experimental observations for varying viscosity ratio ($Pe= 1.4 \times 10^5$). The effect of mesh orientation (parallel or diagonal) is shown, as well as the impact of permeability realisation for the parallel mesh at $M= 20$ (small purple triangles). Simulation error bars represent variations in breakthrough times using a threshold concentration of $c = 0.03 \pm 0.02$. The average experimental result (solid black line) is used to draw the eye to the general trend.}
\label{fig_exp_sim_BT_rec} 
\end{figure}

\begin{figure}[ht]
\centering
\includegraphics[width=0.8\textwidth]{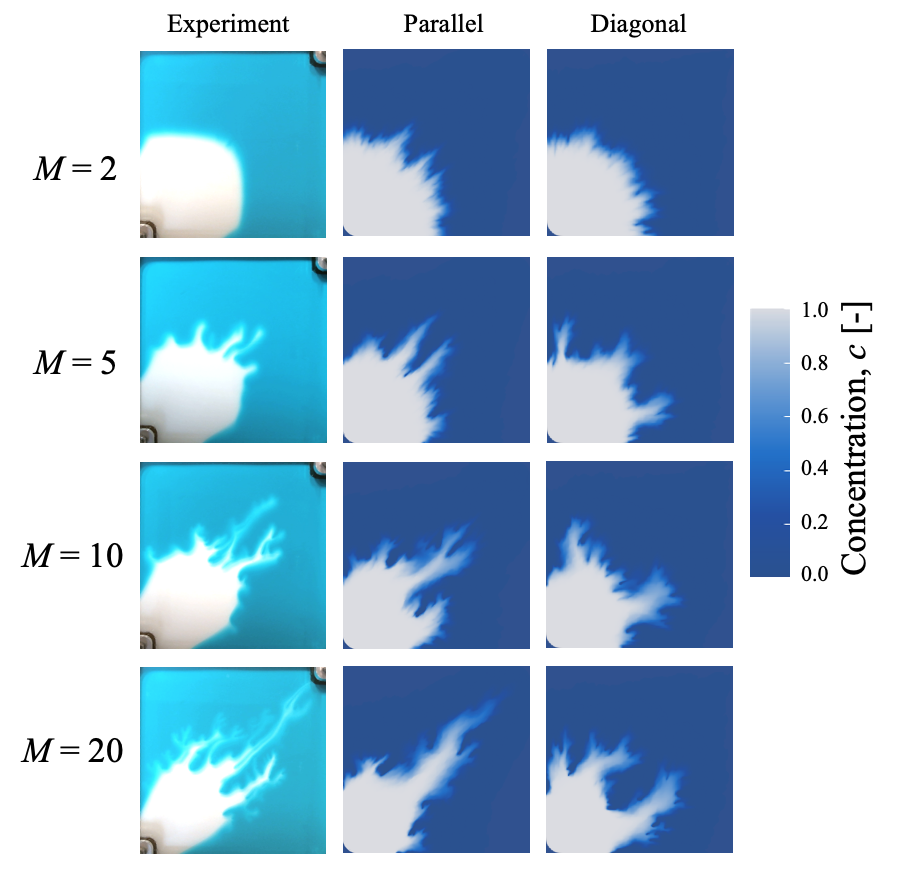}
\caption{Mesh orientation effects (parallel or diagonal) on finger development for different viscosity ratios at fixed $Pe= 1.4 \times 10^5$, at $PVI = 0.25$ against the experimental results.}
\label{fig_exp_sim_mesh_orient} 
\end{figure}

We now discuss the numerical simulation predictions of the experiments, which are shown in Figures \ref{fig_exp_sim_M} to \ref{fig_exp_sim_mesh_orient}, for both parallel and diagonal meshes. For the parallel mesh, shown first in Figure \ref{fig_exp_sim_M}, the fingering patterns predicted by the simulations are qualitatively similar although a) there are more short, fine fingers seen in the low viscosity ratio simulations and b) there is a greater tendency for one finger to dominate the flow in the higher viscosity ratio displacements. Generally, the simulation predicts a slightly later breakthrough time (Figure \ref{fig_exp_sim_BT_rec}) for all viscosity ratios, although the results are within $10\%$ of the experimental results in most cases. There are mesh orientation impacts on both the finger structure (Figure \ref{fig_exp_sim_mesh_orient}), and the corresponding macroscopic measures. When the flow is aligned with the mesh in the parallel case, fingers initiate preferentially along the inlet-outlet flow path, as opposed to the diagonal mesh case where there is more side branching. The breakthrough time of the parallel cases reduces more quickly with $M$ than the experimental results, and is more in line with the experiments at higher $M$ - at higher mobility ratios a single larger finger dominates at later times, as is seen in the experiments.

Figure \ref{fig_exp_sim_BT_rec} shows the macroscopic predictions using both diagonal and parallel meshes, as well as predictions from different realisations of the underlying permeability field. Here we also show simulation uncertainty by using a threshold in concentration which the outlet must reach for break through to have occurred, $c = 0.03 \pm 0.02$. This highlights a `detection' limit, in a similar manner to the experimental results. It also accounts for numerical errors that can occur when considering breakthrough times as $c>0$. These uncertainties can be seen to be less than those inherent in the mesh orientation, and permeability realization, and are largely within the plot symbols. Generally, we see that  the breakthrough times for diagonal meshes are later than for the parallel mesh, as expected, with the parallel mesh case much closer to the experimental results. Changing the permeability realisation can alter the breakthrough time by $\pm 0.03PV$, similar to the uncertainty in repeat experimental runs. Earlier simulations by \cite{Djabbarov2016} used a nine-point centred numerical approximation;  these provided predictions with less mesh orientation error, however, their simulations did not model physical diffusion or velocity dependent dispersion.  Here,  we can see that the simulation uncertainties largely overlap with the experimental uncertainty, and provide good overall predictions of the macroscopic measures.

\subsection{Varying Peclet number, $Pe$}
\label{sec_results_varying_pe}

\begin{figure}
\subfloat[]{\includegraphics[width=\textwidth]{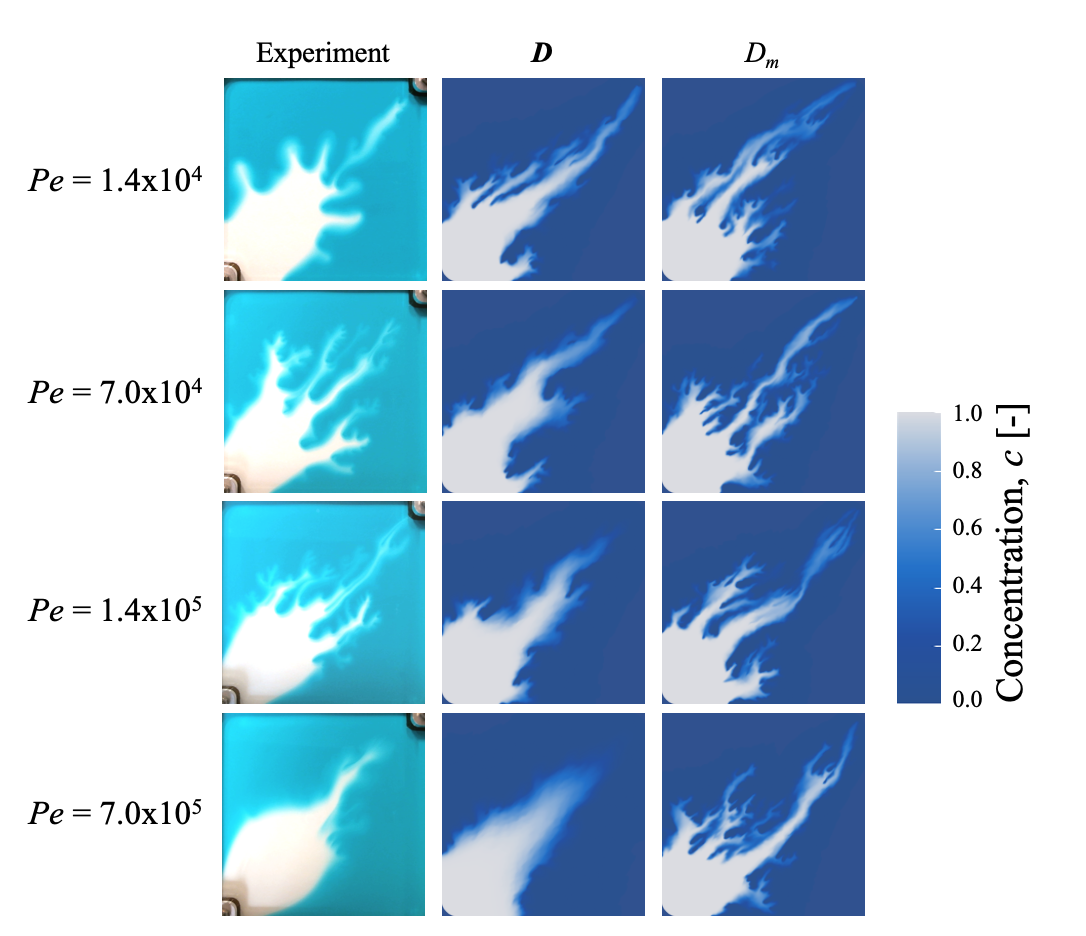}}

\subfloat[]{\includegraphics[width=\textwidth]{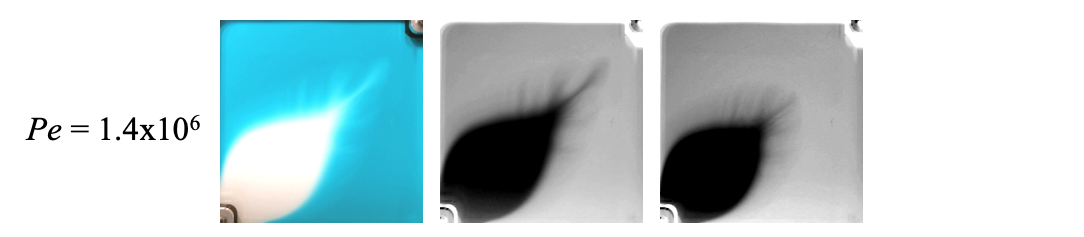}}
\caption{(a) Influence of Peclet number on the fingering pattern for $M = 20$, $t = 0.25 PVI$. First column are experimental results, middle and right columns are simulation results using a parallel mesh – the middle column features dispersion and molecular diffusion, whilst the right column only has molecular diffusion. Note, the highest $Pe$ experiment was not simulated as the experimental fingering pattern was 3D rather than 2D.  (b) High peclet number experimental fingering patterns at $Pe= 1.4 \times 10^6$. In the middle ($t = 0.25 PVI$) and right plots ($t = 0.20 PVI$), the image has been inverted and contrast enhanced to highlight the small wavelength streaks discussed in the main text.}
\label{fig_exp_sim_Pe_plots} 
\end{figure}

The fingering patterns observed in the $M= 20$ experiments at 0.25 PVI for different transverse Peclet numbers (calculated using $D_T = D_m$) are shown in Figure \ref{fig_exp_sim_Pe_plots}, left column with simulation results in the middle and right columns. In these displacements, the longitudinal dispersion (dependent on $v^2$) controls the time at which fingers start to grow, whilst the transverse diffusion (which only depend on molecular diffusivity) and $M$ control the number of fingers and their growth rate. This means that at low $Pe$ transverse diffusion is relatively high and longitudinal dispersion is relatively low so the fingers grow sooner in the displacement, and are able to develop into defined finger structures. As we transition to higher $Pe$ the longitudinal dispersion becomes more important, meaning fingers start growing later from a smeared, more stable interface. 

The fingering pattern in Figure \ref{fig_exp_sim_Pe_plots} changes as $Pe$ is increased from $Pe = 1.4 \times 10^4$ to $Pe = 7 \times 10^5$ and  $1.4 \times 10^6$, with numerous fine fingers forming along the interface between the water and the glycerol mixture. For the highest $Pe$ case, we show enhanced experimental images in Figure \ref{fig_exp_sim_Pe_plots}b, middle and right for $t = 0.25$ and 0.2$PVI$, respectively. This highlights the small, fine fingers along the interface which appear across the gap-width of the cell.  The fingering pattern for $Pe = 7 \times 10^5$ is a combination of one larger scale finger, similar in width to those seen in the lower Peclet number experiments, and very fine fingers, as seen for $Pe = 1.4 \times 10^6$. We infer that this shows a transition from 2-dimensional fingering (where fingers fill the gap between the glass plates) to 3-dimensional fingering where the finger widths are smaller than the plate separation. This transition from 2D to 3D fingering in a Hele Shaw cell has been investigated in more detail by \cite{Videbek2019}. They observed a sudden transition from one type of fingering to the other at a critical Peclet number, where they defined the Peclet number as:
\begin{equation}
Pe_{VN} = \frac{Vh}{D}
\end{equation}
where $V=Q/C \pi r h$ and $C=1/2$ or $2$ depending upon whether a quarter five spot or radial injection is considered.  However, they do not specify the value of interface radius, $r$, where this critical value was calculated. From our results with $r=10$cm, the critical value of $Pe_{VN} $ can be estimated as $\approx$1800 compared with the value of $\approx$1000 reported by Videbæk and Nagel (2019). The critical onset is the same order of magnitude in both cases, however, uncertainty in the radius definition precludes further quantitative comparison. In the earlier work by \cite{Petitjeans1999}, they found $3D$ effects for high viscosity ratio cases, even at relatively low $Pe$ numbers. They used the Atwood number, $At = (\mu_2 - \mu_1)/(\mu_1 + \mu_2)$ to describe the flow regime where a significant portion of the displaced fluid is left trailing on the walls of the cell, at $At > 0.5$. In this regime, there exists an effective interfacial tension between the fluids, forming different fluid-fluid curvatures in the plane of the cell, similar to the immiscible case. The fingers can be impacted by gravity, depending on the density contrast, and rotational flow altering the shape and propagation through the cell. For the experiments here, $At> 0.95$ for $M = 20$, and the density of the displacing and displaced fluids are $\rho_1 = 1$g/cm and $\rho_2 = 1.15$g/cm$^3$, respectively. This suggests 3D effects could be occurring, and the invading finger could buoyantly lift in the plane of cell. 

\begin{figure}[hb!]
\hspace{-35pt}
\subfloat[]{\includegraphics[width=0.65\textwidth]{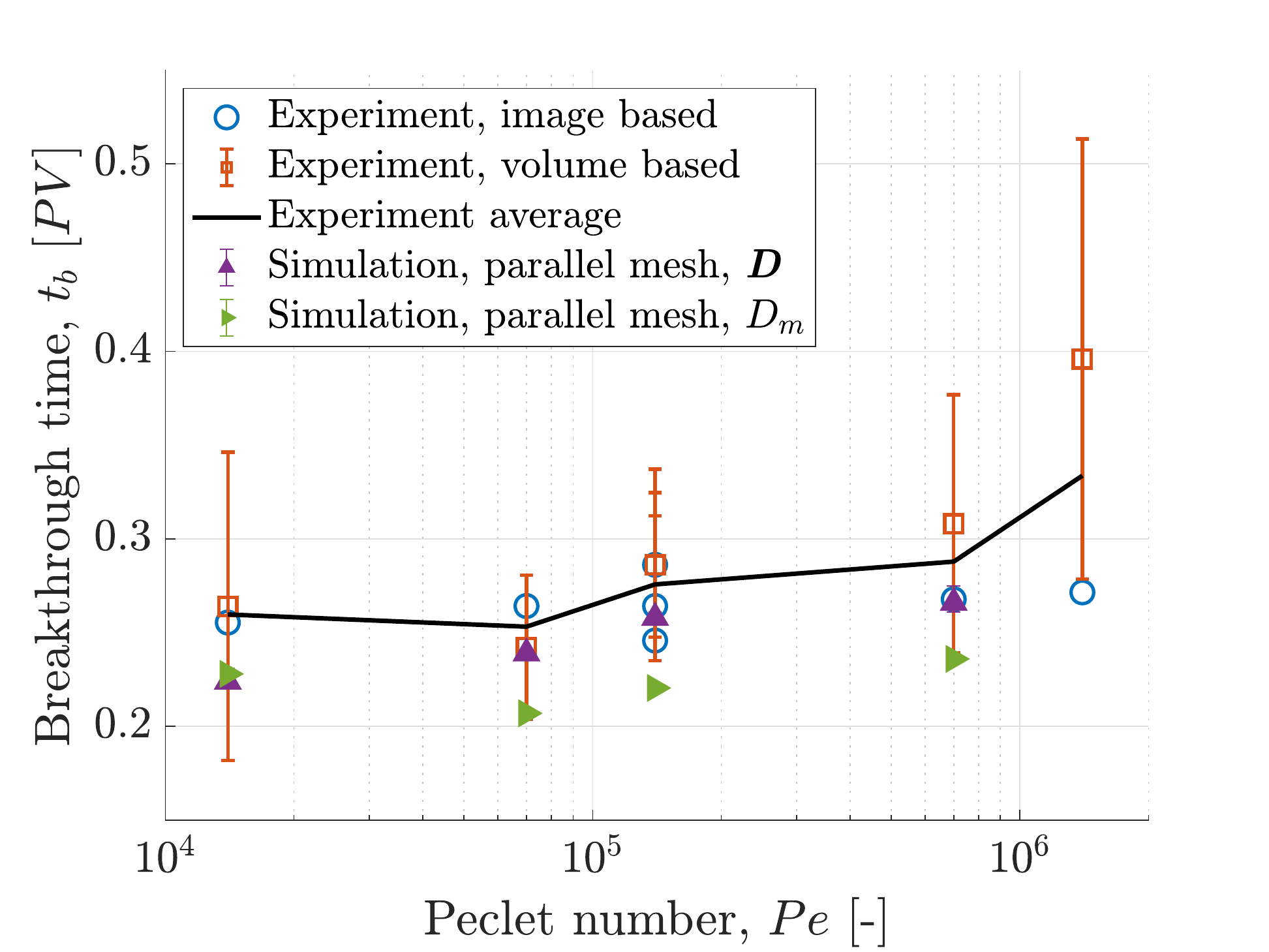}}
\hspace{-20pt}
\subfloat[]{\includegraphics[width=0.65\textwidth]{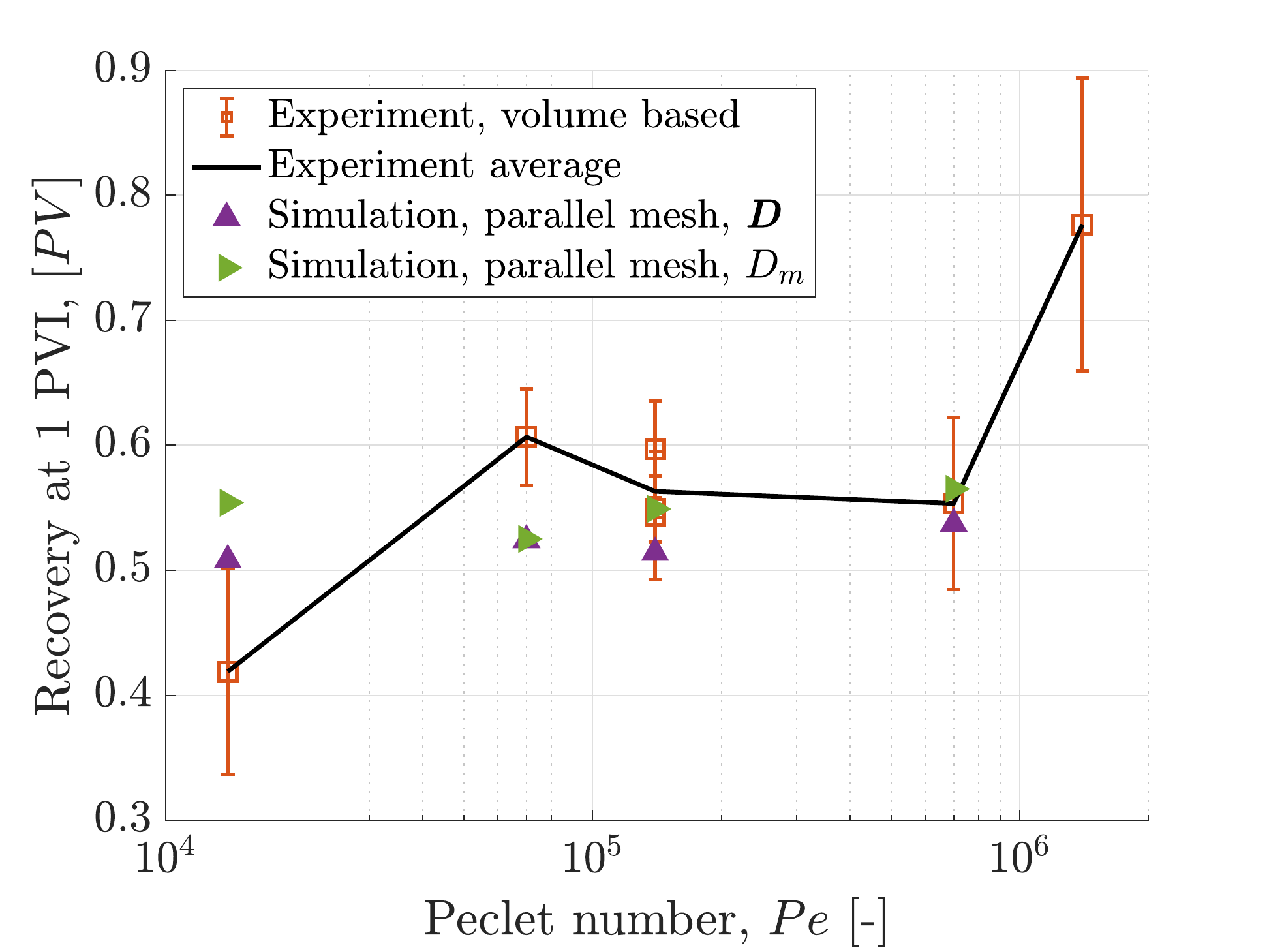}}
\caption{Numerical simulation predictions of (a) breakthrough time (in pore-volumes, $PV$) and (b) recovery after 1 pore-volume injected $PVI$, with experimental observations for varying Peclet number ($M=20$). No simulation predictions are shown for the highest Peclet number because the simulations were 2D. Simulation error bars represent variations in breakthrough times using a threshold concentration of $c = 0.03 \pm 0.01$. The average experimental result (solid black line) is used to draw the eye to the general trend. Note, the volumetric results at $Pe= 1.4 \times 10^6$ have large experimental uncertainty, potentially due to 3D effects which lead to small concentrations obfuscating breakthrough time prediction.}
\label{fig_exp_sim_bt_rec_Pe} 
\end{figure}

Simulation predictions at varying $Pe$ are shown in Figure \ref{fig_exp_sim_Pe_plots}a middle and right columns, with macroscopic predictions shown in Figure \ref{fig_exp_sim_bt_rec_Pe}. We show two model results, one using the full dispersion tensor $\bm{D}$ in Equation (\ref{eqn_dispersion_tensor}) and (\ref{eqn_dispersion}), the other using purely molecular diffusion, e.g. $D_L = D_T = D_m$. In general, we see that the two models with dispersion and pure diffusion, respectively, somewhat bound the experimental finger evolution. The model with pure molecular diffusion predicts a more unstable finger regime at the (high) Peclet numbers considered here, with the results largely invariant to changes in $Pe$ at this level. All $Pe$ regimes have similar wavelength and number of fingers. This is further reflected in the breakthrough time and recovery, which is essentially constant for the molecular diffusion model. This shows the influence (or lack there-of) of velocity dependent longitudinal dispersion on the time at which fingers grow

The model with dispersion predicts a stabilisation of the interface as $Pe$ is increased. This is similar to the numerical results of \cite{Petitjeans1999}, but at much higher $Pe$ here. This enhanced dispersion both increases the breakthrough time and recovery (sweep efficiency) as $Pe$ is increased, which is in-line with the experiment results. However, the mechanisms for the enhancements are somewhat different; in the experiment, it is clear that 3D fingering is occurring, with fingers forming in the plane of the cell-thickness, which cause a greater (areal) sweep of the cell. In the simulation, the interface in the cell x-y plane is dispersed and smoothed (somewhat a result of numerical diffusion), somewhat due to a delayed breakthrough and increased sweep efficiency. It is clear from the experiments that there is greater instability and more fingers than the dispersive model, but the resulting macroscopic measures are very similar in each case due to the 3D effects.

In both simulation models, the fingering is somewhat sharper than the experiments, and finer fingers are not well captured through their full evolution. The simulations use a fine mesh resolution of 1mm, however some of the experimental fingers approach this size and may not be well resolved. There may also be 3D effects, even in the lower $Pe$ cases, as noted in \cite{Petitjeans1999} as the viscosity ratio is still high. The Atwood number, $At > 0.95$ for the $M = 20$ cases here, meaning a significant fraction of the cell thickness may be left uninvaded, changing the apparent `sharpness' of the fingers themselves. With this behaviour, the flow may diverge from Poiseuille behaviour, meaning the velocity dependence in the dispersion model may not be quadratic. We also note that we assume transverse dispersion is equal to molecular diffusion here; this too could be dependent on velocity and have impacts on the interactions between fingers, potentially enabling more intricate finger structures as seen in the experiments here (\cite{Ghesmat2007}). 

In this section, we have shown various model predictions of the miscible fingering in a quarter five-spot Hele-Shaw configuration. The models show that we can predict macroscopic behaviour relatively well, and trends in $Pe$ and $M$ dependence. However, finger growth mechanisms are still mismatched with the experiment. The experimental dataset herein provides a useful starting point to explore extensions to existing theory to cover the full spectrum of $M$ and $Pe$ effects observed in experiments, with time-lapse images also allowing the onset of fingering (and subsequent evolution) to be rigorously explored.

\section{Conclusion}
\label{sec_conclusions}

A series of well characterised experiments investigating miscible viscous fingering in radial flow using a Hele Shaw cell have been performed and their results compared with the predictions of high-resolution finite difference simulations using various mesh and model choices, across a range of Peclet numbers and viscosity ratio. We performed repeat experiments in opposite flow directions, and in the same flow direction, to quantify the underlying uncertainty, which is paramount for assessing model conformance. 

The dynamics of the fingering patterns seen in the Hele-Shaw experiments were consistent with previous experiments. For modest Peclet numbers, the number of fingers, their growth rate and the level of side branching increases with the viscosity ratio between the injected and displaced fluids. In addition, we see a transition from 2D to 3D fingers as Peclet number increases. This transition occurs at the same (order of magnitude) critical Peclet number as seen by \cite{Videbek2019}.

Our simulations generally provided good predictions of the macroscopic behaviour in terms of breakthrough time and recovery as a function of viscosity ratio, for intermediate Peclet numbers. Mesh orientation effects were observed, with parallel and diagonal orthogonal meshes producing results which effectively bounded the behaviour seen in the experiments. The difference in mesh behaviour may stem from the initiation of the fingers from the underlying permeability structure - the orientation of the mesh impacts the primary direction with which fingers first initiate. The experiments show that although macroscopic behaviour is repeatable between experiments, the specific fine-scale fingers that form are not, although they have overall globally similar wavelength and number. The difference appears to stem from the inlet of the fluid, and the initial interaction of the stable interface with slightly different parts of the `imperfect' cell.

We used two models to investigate the Peclet number behaviour; one with transverse dispersion dependent on the squared interface velocity, and the other using pure molecular diffusion. We found the dispersive model generally predicted the macroscopic breakthrough and recovery behavior across the $Pe$ range well, although the specific fine finger growth did not match the experiments. The simulations had a more dispersive interface in the plane of the cell than the experiments. The molecular diffusion model on the other hand did predict the formation of smaller fingers, but showed little $Pe$ dependency. Both models predicted sharper fingers than were found in the experiments. In the experiments, at $M=20$ for the highest Peclet numbers, 3D effects in the plane of the cell thickness appeared considerable (and potentially for lower $Pe$ cases, although direct measurement with the current setup was precluded), altering the finger shape and evolution. Further work should model the full 3D problem with various dispersion models with modern, parallelised code to fully explore the range of $M$ and $Pe$ behaviour. 

All the results from these experiments are available in open-access form, see the Data access section below. They represent an ideal benchmark dataset for the future development of models and numerical methods to simulate viscous instabilities.

\section{Acknowledgements}
\label{sec_acknowledgements}
Dr. Wawrzyniec Kostorz is thanked for adding several new features to the MISTRESS code that enabled this study to be performed. 

\section{Data access}
\label{sec_data_access}

Experimental and simulation data associated with this work are hosted on \textcolor{blue}{\href{https://zenodo.org}{zenodo.org}} with DOI: 10.5281/zenodo.5567913. The matlab codes for processing the images, analysing results and generating figures, as well as example simulation datafiles are available on Github at \textcolor{blue}{\href{https://github.com/sci-sjj/MiscibleViscousFingering}{https://github.com/sci-sjj/MiscibleViscousFingering}}. The Mistress simulation code is currently under-development; up to date code is available from the authors on request. 

\section{Statements and Declarations}
The authors declare that they have no known competing financial interests or personal relationships that could have influenced the work in this paper.

\section*{Appendix A.}
\label{sec_appendix_A}

\begin{table}[ht]
\hspace{-20pt}
\begin{tabular}{|c|c|c|c|c|c|c|}
\hline
\textbf{\begin{tabular}[c]{@{}c@{}}Mobility \\ ratio, \\ $M$\end{tabular}} & \textbf{\begin{tabular}[c]{@{}c@{}}Flow \\ rate \\ {[}ml/min{]}\end{tabular}} & \textbf{\begin{tabular}[c]{@{}c@{}}Peclet \\ number, \\ $Pe$\end{tabular}} & \textbf{\begin{tabular}[c]{@{}c@{}}Break\\ through \\ time {[}$PV${]}, \\ Image\end{tabular}} & \textbf{\begin{tabular}[c]{@{}c@{}}Break\\ through \\ time {[}$PV${]}, \\ Volumetric\end{tabular}} & \textbf{\begin{tabular}[c]{@{}c@{}}Recovery \\ at 1$PVI$ {[}$PV${]},\\ Volumetric\end{tabular}} & \textbf{\begin{tabular}[c]{@{}c@{}}Error {[}$PV${]},\\ Volumetric\end{tabular}} \\ \hline
2                                                                          & 1.0                                                                           & $1.4 \times 10^5$                                                          & 0.448                                                                                       & 0.484                                                                                            & 0.706                                                                                       & 0.076                                                                         \\ \hline
5                                                                          & 1.0                                                                           & $1.4 \times 10^5$                                                          & -                                                                                           & 0.396                                                                                            & 0.610                                                                                       & 0.052                                                                         \\ \hline
5                                                                          & 1.0                                                                           & $1.4 \times 10^5$                                                          & 0.341                                                                                       & 0.374                                                                                            & 0.699                                                                                       & 0.003                                                                         \\ \hline
10                                                                         & 1.0                                                                           & $1.4 \times 10^5$                                                          & 0.308                                                                                       & 0.352                                                                                            & 0.606                                                                                       & 0.021                                                                         \\ \hline
10                                                                         & 1.0                                                                           & $1.4 \times 10^5$                                                          & 0.286                                                                                       & 0.330                                                                                            & 0.552                                                                                       & 0.067                                                                         \\ \hline
20                                                                         & 1.0                                                                           & $1.4 \times 10^5$                                                          & 0.286                                                                                       & 0.286                                                                                            & 0.549                                                                                       & 0.026                                                                         \\ \hline
20                                                                         & 1.0                                                                           & $1.4 \times 10^5$                                                          & 0.264                                                                                       & 0.286                                                                                            & 0.543                                                                                       & 0.051                                                                         \\ \hline
20                                                                         & 1.0                                                                           & $1.4 \times 10^5$                                                          & 0.246                                                                                       & 0.286                                                                                            & 0.597                                                                                       & 0.039                                                                         \\ \hline
20                                                                         & 0.1                                                                           & $1.4 \times 10^4$                                                          & 0.255                                                                                       & 0.264                                                                                            & 0.419                                                                                       & 0.082                                                                         \\ \hline
20                                                                         & 0.5                                                                           & $7.0 \times 10^4$                                                          & 0.264                                                                                       & 0.242                                                                                            & 0.607                                                                                       & 0.039                                                                         \\ \hline
20                                                                         & 5.0                                                                           & $7.0 \times 10^5$                                                          & 0.268                                                                                       & 0.308                                                                                            & 0.553                                                                                       & 0.069                                                                         \\ \hline
20                                                                         & 10.0                                                                          & $1.4 \times 10^6$                                                          & 0.272                                                                                       & 0.396                                                                                            & 0.776                                                                                       & 0.117                                                                         \\ \hline
\end{tabular}
\caption{\label{tab_exp_results}Tabulated experimental results. Error [$PV$], Volumetric refers to the difference in displaced volume and displacing volume at the estimated breakthrough point, as described in the text. Quoted $M$ values have relative uncertainties of $\pm$10\% to capture the range of experimental measurements.}
\end{table}

\end{document}